# Contactless Transport and Mixing of Liquids on Self-Sustained Sublimating Coatings

*Athanasios Milionis[‡,†], Carlo Antonini[‡,†,§], Stefan Jung[†], Anders Nelson[†], Thomas M. Schutzius[†]*

*and Dimos Poulikakos*[\*,†]

[†]Laboratory of Thermodynamics in Emerging Technologies, Institute of Energy Technology,

Department of Mechanical and Process Engineering, ETH Zürich, 8092 Zürich, Switzerland.

[§]Empa, Swiss Federal Laboratories for Materials Science and Technology,

Functional Cellulose Materials, Überlandstrasse 129, CH-8600 Dübendorf, Switzerland

**ABSTRACT:** The controlled handling of liquids and colloidal suspensions as they interact with surfaces, targeting a broad palette of related functionalities, is of great importance in science, technology, and nature. When small liquid volumes (drops of the order of µl or nl) need to be processed in microfluidic devices, contamination on the solid/liquid interface and loss of liquid due to adhesion on the transport channels are two very common problems that can significantly alter the process outcome, e.g. the chemical reaction efficiency, or the purity and the final concentrations of a suspension. It is therefore no surprise that both levitation and minimal contact transport methods—including non-wetting surfaces—have been developed to minimize the interactions between liquids and surfaces. Here we demonstrate contactless surface levitation and transport of liquid drops, realized by harnessing and sustaining the natural sublimation of a solid carbon dioxide-coated substrate to generate a continuous supporting vapor layer. The capability and limitations of this technique in handling liquids of extreme surface tension and kinematic viscosity values are investigated both experimentally and theoretically. The sublimating coating is capable of repelling many viscous and low-surface tension liquids, colloidal suspensions, and non-Newtonian fluids as well, displaying outstanding omniphobic properties. Finally, we demonstrate how sublimation can be used for liquid transport, mixing and drop coalescence, with a sublimating layer coated on an underlying substrate with prefabricated channels, conferring omniphobicity with a simple physical approach (i.e. phase-change), rather than a chemical one. The independence of the surface levitation principle from material properties, such as electromagnetic, thermal or optical, surface energy, adhesion or mechanical properties renders this method attractive for a wide range of potential applications.

## INTRODUCTION

The capability to handle and manipulate drops offers significant advantages in many applications. In devices such as lab-on-a-chip or other microfluidic manifestations, drop handling enables sample volumes to be significantly reduced, leading to concomitant reductions in cost and time.[1,2] However, understanding how a drop will interact with a solid surface is far from simple, due to the complexity and variety of factors and mechanisms that can intervene at different length scales.[3] Depending on the conditions, a drop on a solid surface may deposit, spread, stick, slide, splash, break-up, partially or completely rebound, levitate, evaporate, and/or solidify.[4,5] From a practical standpoint, contact with a solid surface also leads to sample contamination while at small length scales introduces difficulties in controlling mobility, due to strong capillary effects.[6]

Different strategies can be typically found in the literature as possible routes for facilitating drop motion on surfaces. The first strategy, which involves ongoing research during the last two decades, is the development of bioinspired superhydrophobic surfaces (typically identified by the arbitrary threshold of contact angles $\theta > 150º$, or alternatively with receding contact angles $\theta_R > 135º$, as based on drop mobility[7] and drop impact[8] experimental study) accompanied with low contact angle hysteresis ($\Delta\theta < 10º$).[9,10] The unique functionality of superhydrophobic surfaces lies in their micro-/nano-textured surface topography, which combined with appropriate surface chemistry, e.g., hydrophobic polymers or micro-/nano-particles, impart extreme water repellency. During the last decade, omniphobic surfaces with re-entrant surface curvature were designed and developed to repel a wider selection of low-surface tension liquids, in addition to water.[11-15] More recently, a second strategy in liquid repellency based on lubricant impregnated porous surfaces (commonly abbreviated as SLIPS or LIS) has been developed: such surfaces feature lower wetting angles, but also very low values of contact angle hysteresis owing to the defect-free nature of the lubricant-air interface.[15-20]

There are drawbacks and limitations in all the aforementioned approaches. Contact between the liquid and the substrate, although minimized, always occurs during the transportation of the liquid, which may be critical for application where the sample needs to be preserved from contamination. Superhydrophobic and omniphobic surfaces[9–15] rely on the presence of vapor pockets at the solid-liquid interface to minimize the contact (the so called Cassie-Baxter state), but under varying pressure[21] and during condensation/frost[22,23] formation processes, liquid may penetrate in the surface texture, displacing vapor pockets, so that the drop loses its mobility and sticks to the surface (Wenzel state). Surface durability on these type of surfaces is also poor in most of the cases.[24] In the case of lubricant impregnated porous surfaces, applications are limited due to some of the above reasons (contamination and mechanical robustness), as well as technological issues, such as the loss of the infused lubricating liquid over time,[25] the penetration and chemical interaction of the lubrication medium with the handled liquid, and temperature sensitivity. Therefore, new passive solutions alleviating some of the above issues are highly desirable, since they would be appropriate for a range of applications – such as stringent requirements of non-contamination - where existing strategies are not working. The focus of this work is not only to minimize, but to eliminate the contact between the surface and the handled liquid drops, while maintaining the ability to control drop motion.

In 1756, J. G. Leidenfrost discovered that a water drop can float on hot substrate, above 300°C, due to rapid drop evaporation at the interface. The principle was also extended to allow self-propulsion of a sublimating body on a metal ratchet.[26,27] More recently, an inverted Leidenfrost effect was demonstrated during condensation of water drops on frozen liquid layers.[28] Although fascinating, several factors limit applications for the classical Leidenfrost levitation mechanism, such as drop evaporation, drop temperature control, and dependence on liquid properties, e.g. boiling temperature. Due to the aforementioned limitations, the integration of this physical concept to lab-on-a-chip and microfluidic devices where liquids have to be processed with extreme precision is not recommended and alternative contactless approaches have to be followed.

On the other hand, only a few studies have reported contactless handling of matter on the basis of electromagnetic,[29,30] acoustic[31,32] and sublimation principles,[33,34] the first two being active techniques. The idea underpinning the strategy proposed here is to extend our previous basic findings of drop surface levitation phenomena on sublimating surfaces,[33,34] by examining the range of application-relevant liquids for which the surface levitation process can be applied, and realize the proof-of-concept on a channel system that can perform contactless transport and mixing/coalescence of various liquids. In our previous works,[33,34] we have already reported the remarkable non-wetting properties of solid carbon dioxide ($CO_2$), commonly known as dry ice, in repelling both water and highly viscous glycol drops during impact experiments: In particular, we observed that water drops rebound after impact, as typically observed in Leidenfrost boiling conditions[35] and on superhydrophobic surfaces,[36,37] and that a different mechanism of tumbling rebound applies even to highly viscous liquid drops, in case of non-axisymmetric impact conditions.

Here we realized an experimental method and apparatus for obtaining self-sustained perfectly non-wetting $CO_2$ coatings on aluminum channels for manipulating a wide variety of liquids in a contactless manner. The self-regeneration and sustenance of the coatings occurs due to continuous supply of $CO_2$ vapor, which desublimates (i.e. deposits, by vapor-to-solid phase-change) on a cooled substrate with prefabricated open channels. As such, on one hand the $CO_2$ sublimates under the handled liquid drops, allowing surface levitation. On the other hand, the $CO_2$ coating is continuously restored by desublimation on the (cold) solid substrate (see Figure 1 in the results and discussion). To explore the potential of sublimating surfaces and coatings, we studied the effect of liquid kinematic viscosity and surface tension on omniphobicity achieved by surface levitation. After carefully mapping the liquids used in this work, we identified two conditions related to the liquid kinematic viscosity and surface tension values that need to be satisfied for achieving surface levitation. Then, examples of drop mixing/coalescence were demonstrated, using different types of Newtonian and non-Newtonian liquids, such as colloidal and polymer suspensions, by monitoring the coalescence process with a high speed camera. The contactless process of combining small volumes of liquids in the surface levitation regime is expected to find potential applications in high-precision lab-on-a-chip and other microfluidic devices.

## RESULTS AND DISCUSSION

**Surface levitation mechanism.** Motivated by the challenge to create surfaces that can fully repel liquids, and overcome the limitation of liquid impalement, particularly critical in dynamic conditions, we identified sublimating surfaces as an ideal candidate for their exceptional omniphobic properties. The concept is applicable to most liquids, as can be seen by the images of different types of liquid drops (water, glycerol, silicone oil and decane) deposited on a sublimating substrate in Figure 1a. When a room temperature liquid drop is deposited on a sublimating surface, a carbon dioxide vapor layer is instantaneously formed at the drop-substrate interface due to sublimation of the substrate, as schematically illustrated in Figure 1b, so that drops can levitate on the surface, in a perfect non-wetting state without freezing for the time required for the coalescence process (typically in the order of hundreds of milliseconds). The vapor layer plays essentially a double role, both as an air cushion that levitates the drop, and as insulating layer to prevent freezing. Indeed, the conductive heat transfer between the liquid (room temperature ≈ 20 ºC) and the substrate (≈ -80 ºC) through the vapor layer is sufficient to promote substrate sublimation, which is the key factor for drop surface levitation, while at the same time low enough, so that the liquid drops can rebound in case of impact, or in any case, stay in the liquid phase without freezing for a time period in the order of seconds. Since the drop does not contact the solid substrate, we could define for the surface a virtual contact angle of 180° and a contact angle hysteresis of 0° (see

Figure 1b) – although strictly speaking contact angles cannot be defined, since there is no contact - conferring to the surface its perfect non-wetting condition.

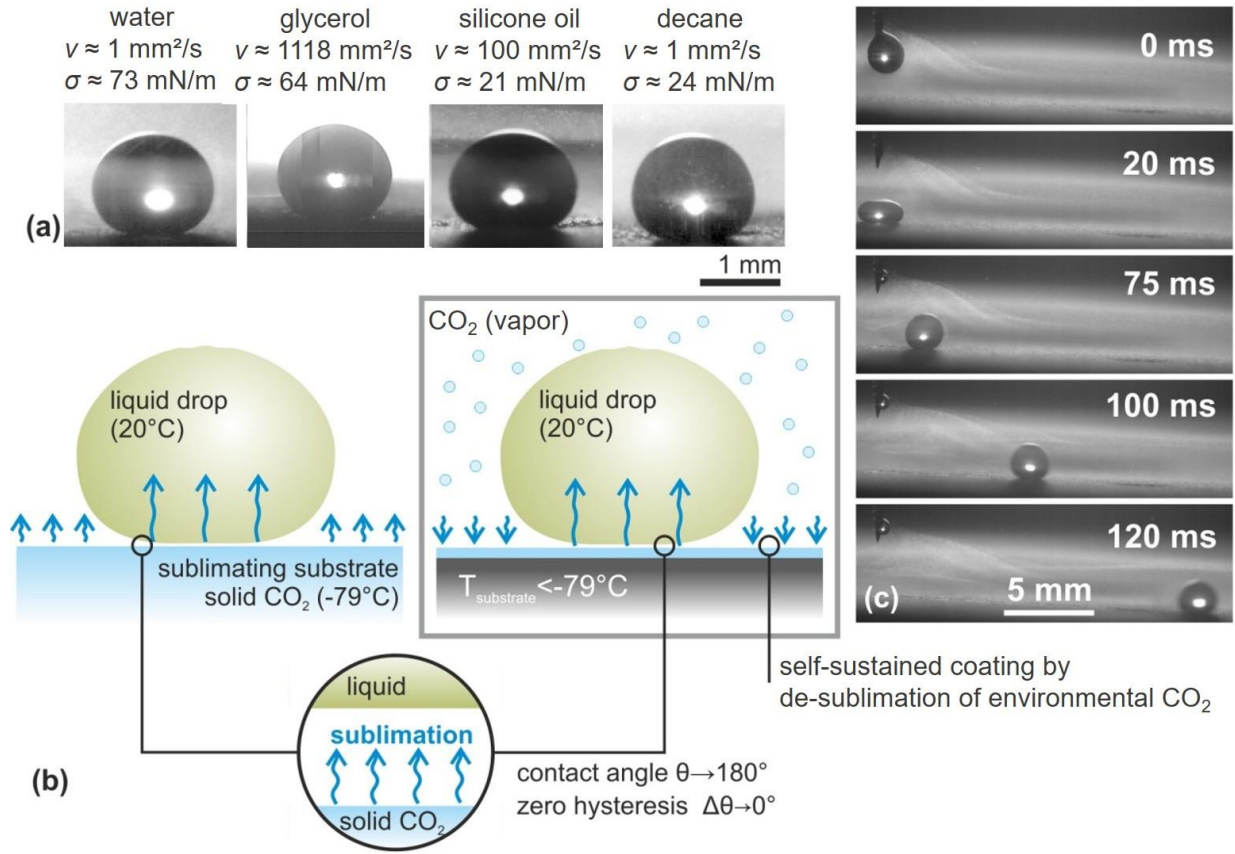

*Figure 1.* (a) Images of a selection of various liquid drops gently deposited on a $CO_2$ sublimating surface. The liquid name, the kinematic viscosity (v) and surface tension values (σ) are provided above the drop images. (b) Schematic of the contactless interaction between a liquid drop and the sublimating surface (substrate and coating). The coating can be continuously self-sustained due to constant desublimation on the cold substrate of the $CO_2$ vapor in a closed chamber, at room temperature. (c) Image sequence of a water drop rapidly rolling down a $CO_2$ sublimating substrate. The tilt in this case is 2°. However, even for the smallest tilt angles, practically close to 0°, liquid drops could also roll down the sublimating substrates.

This is also demonstrated by the image sequence in Figure 1c where a water drop is dispensed on a sublimating surface tilted at 2°: the water drop easily rolls off the surface. Rolling off was observed in almost all the cases, including smaller drops, below 1 mm diameter (reduced gravitational force). Even on perfectly planar substrates, it was difficult to deposit and keep in place liquid drops on the sublimating coatings (see Movie M1), due to the absence of capillary adhesion between the drop and the substrate.

**Surface levitation *vs.* drop break-up.** To investigate the conditions for omniphobic behavior of the sublimating surfaces for various liquids, a wide selection of liquids with different characteristics was tested, with surface tension ranging from 11.9 to 72.8 mN/m and kinematic viscosity ranging from 0.3 to 1117 mm$^2$/s. It is important to understand which physical parameters influence the phenomenon of surface levitation via dry ice sublimation and evaluate them. The $CO_2$ vapor generated at the sublimating substrate interface forms a layer that can sustain the handled liquid, while flowing radially outwards: this leads to the surface levitation regime. However, this regime is not reached with all liquids: in some cases, a "drop break-up" was also observed. Movies M1 and M2 show two representative examples for the two observed different regimes. In Movie M1 a silicone oil drop (surface tension 20.6 mN/m and kinematic viscosity 20 mm$^2$/s) is levitated on the $CO_2$ substrate in a contactless manner after deposition, whereas in Movie M2 drop break-up due to bubbling can be observed for perfluorohexane (surface tension 11.9 mN/m and kinematic viscosity 0.4 mm$^2$/s). In this second case, the $CO_2$ vapor breaks the drop-vapor interface, forming bubbles inside the drop. As such, the $CO_2$ vapor penetrates the drop, causing the observable bubbling (which gives the false impression of a liquid being in boiling state), instead of escaping from the periphery of the drop. The movement of the vapor bubbles inside the liquid is dependent on the liquid kinematic viscosity, as we will discuss below.

The physical quantities that will resist $CO_2$ vapor penetration are the liquid surface tension ($\sigma$) and viscosity ($\mu$), while gravity (high drop density, $\rho$) will tend to favor drop bubbling.

High $\sigma$ implies that the liquid surface will be more resistive to deformation induced by the vapor pressure at the vapor-liquid interface. Also, high liquid viscosity $\mu$ slows down the interface deformation and delays the propagation of vapor bubbles eventually penetrating into the drop, so that bubbles propagation in the liquid can be retarded on a time scale larger than the milliseconds required for liquid coalescence and transport. On the other hand, a denser drop will favor bubbling since it will increase the hydrostatic pressure applied on the underlying $CO_2$ vapor layer. Moreover, for a fixed drop diameter, increased $\rho$ will reduce the thickness of the $CO_2$ vapor layer separating the drop from the dry ice coating, making it harder for the $CO_2$ vapor to escape through the drop perimeter. To account for both viscosity and density effects, it is thus appropriate to analyze the different regimes using the kinematic viscosity, $v=\mu/\rho$.

**Regime mapping for various liquids on the sublimating surface.** To evaluate the efficiency of the sublimating coatings to repel different liquids, approximately 10-15 μl size drops (2.7-3.1 mm diameter) of more than 50 different liquids and liquid mixtures were deposited on slightly inclined dry ice blocks (tilt angle ≈ 2º) to observe the rolling motion of the drops. These tests were performed for convenience using single-use dry ice blocks. The size of the drops was of the same order of magnitude with their corresponding capillary lengths ($l_c$). Although the diameter effect on levitation was not explicitly explored on the sublimating surface, in a few tests we could observe that smaller drops, generated by break-up, behaved similarly as larger drops with $D/l_c \sim 1$. For example, in the case of isopropyl alcohol ($\sigma$ = 23.7 N/m and $v$ = 2.6 mm$^2$/s) – see Movie M3 – upon drop break-up of a millimetric drop, the generated smaller drops (with diameter 5 to 10 times smaller, which appear left of the main drop during the tenth second in the movie) feature the same bubbling behavior on the sublimating surface. For this reason, we have focused more on the kinematic viscosity and the surface tension as the main parameters to be investigated.

Figure 2 shows a graph that maps the outcomes of drop-$CO_2$ surface contact for many different liquids, on the basis of their surface tension $\sigma$ and kinematic viscosity $v$; surface tension values ranged from 11.9 (perfluorohexane) to 72.8 mN/m (water) and viscosity values ranged from 0.45 (hexane) to 1118 mm$^2$/s (glycerol). Each marker in Figure 2 corresponds to a given liquid with specific $\sigma$ and $v$ values at room temperature. Green closed symbols represent liquids in the surface levitation regime, whereas red open symbols represent liquids in the "break-up" regime. Symbol shapes correspond to different classes of liquids: triangles highlight various water/glycerol mixtures (all in surface levitation); squares highlight silicone oils (see also Table 1); and diamonds denote the chemical group of alkanes (see also Table 2). Most of the liquids in the "break-up" regime are concentrated at the bottom left part of the graph, corresponding to both low surface tension and kinematic viscosity values. Indeed, two threshold conditions can be identified, for a liquid to levitate on the sublimating surfaces. These are: surface tension $\sigma > \sigma_C \approx 20$ mN/m and kinematic viscosity $v > v_C \approx 0.9$ mm/s (non-grey areas of Figure 2). Liquids that do not satisfy these two necessary (but not sufficient) conditions belong to the "break-up" regime. Only two liquids (out of more than 50), i.e. ethanol and isopropyl alcohol, which belong to the chemical family of monohydric alcohols together with methanol (marked with the symbol ∾ in Figure 2), experience break-up, despite having $\sigma$ and $v$ values slightly above, but very close to, the thresholds. However, the overall picture provided by results illustrated in Figure 2 is clear, and the two regimes can be well-predicted for the vast majority of the liquids on the basis of surface tension $\sigma$ and kinematic viscosity $v$.

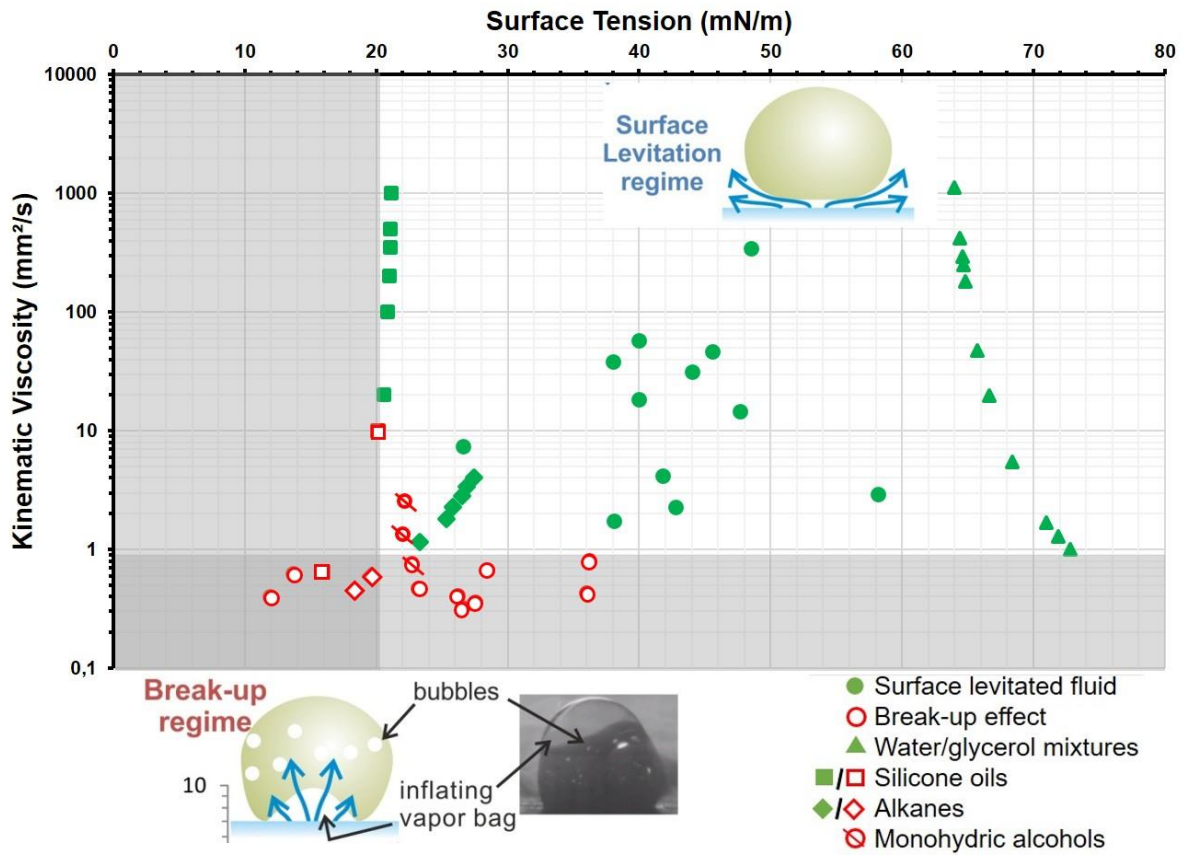

*Figure 2. Regime map of multiple liquid behavior on CO$_2$ sublimating surfaces. Each symbol represents a liquid with a given surface tension and kinematic viscosity (values at room temperature, at which the drop is deposited on the surface). Green closed symbols correspond to liquids in the "surface levitation" regime, whereas red open symbols to the "break-up" regime. Gray regions highlight the region of break-up regime. For all the liquids approximately 10-15 μl size drops (2.7-3.1 mm diameter) were used.*

**Determining the combined role of surface tension and kinematic viscosity on levitation.** The stability of the air/water interfaces has been already studied in the case of the Leidenfrost effect and an inverse Rayleigh-Taylor instability where vapor film tends to rise because of Archimedes' thrust, has been used to explain the interface instability.[38,39] These instabilities take effect when the drop diameter exceeds the critical value $d^*$ which can be estimated by multiplying the capillary length, $l_c = \sqrt{\sigma/\rho g}$, by a factor of 7.6: i.e. $d > d^* = 7.6 l_c$, with $\sigma$ and $\rho$ being the surface tension and density of a given liquid respectively, and g = 9.81 m/s$^2$). For the liquids investigated in the present study, the capillary length is restricted to the range of 0.85 < $l_c$ < 2.72 mm, where the extreme values correspond to perfluorohexane and water (see Movies M1 and M2), and drop diameters range from 2.7 to 3.1 mm as mentioned earlier. For example, perfluorohexane, which has the lowest capillary length ($l_c$ = 0.85 mm) and for which break-up was observed, should become unstable only for drop diameters exceeding the critical value $d^*$ = 6.46 mm, according to the inverse Rayleigh-Taylor instability criterion, thus for drop diameters significantly greater compared the drop diameters used here. As such, drop break-up cannot be explained by the criterion based on the inverse Rayleigh-Taylor instability developed for Leidenfrost boiling.

To perform an in-depth investigation of the surface tension and kinematic viscosity effects in liquid levitation, three different groups of liquids were specifically tested, namely: (i) water/glycerol mixtures, (ii) silicone oils and (iii) alkanes. The first two groups exhibit a relatively constant surface tension, with the possibility to change kinematic viscosity by three orders of magnitude. The water/glycerol mixtures, depending on the water-to-glycerol ratio, exhibit kinematic viscosity values from 1 to 1118 mm$^2$/s, and surface tension values within the narrow range of 64 to 73 mN/m. Similarly, silicone oils exhibit kinematic viscosities from 0.65 to 1000 mm$^2$/s, with surface tension values restrained to a narrow range (16 to 21 mN/m), which are significantly lower compared to water/glycerol mixtures. As shown in Figure 1a and 2, both water and glycerol could be repelled successfully by the sublimating substrates; similar behavior was observed with their varying ratio mixtures for their entire range of kinematic viscosities (liquids with triangle marker in Figure 2). The surface levitation behavior is consistent with our previous observations of liquid drop rebound during drop impact events.[33] The scenario is different for silicone oils, exhibiting similar kinematic viscosities but three times lower surface tension (square symbols in Figure 2 – see also numerical values in Table 1). As it can be observed, the silicone oils with the two lower kinematic viscosity values (0.65 and 10 mm$^2$/s) and lower surface tension (up to ~ 20 mN/m) could not be levitated by the dry ice sublimating surface. On the other hand, silicone oils with high kinematic viscosity values, 10 mm$^2$/s and above, and $\sigma > \sigma_C \approx 20$ mN/m could successfully levitate.

**Table 1.** Levitation outcome of silicone oils with varying viscosities on the sublimating surfaces. The kinematic viscosity and surface tension values were obtained from the literature.

| Kinematic Viscosity (mm$^2$/s) | Surface Tension (mN/m) | Levitation outcome |
|---|---|---|
| 0.65 | 15.9 | Break-up |
| 10 | 20.1 | Break-up |
| 20 | 20.6 | Surface levitation |
| 100 | 20.9 | Surface levitation |
| 200 | 21.0 | Surface levitation |
| 350 | 21.1 | Surface levitation |
| 500 | 21.1 | Surface levitation |
| 1000 | 21.2 | Surface levitation |

To underpin the role of $\sigma$ in resisting bubble formation, it is instructive to compare liquids of relatively similar viscosity (varying in a narrower range) and different surface tension: With silicone oils it was possible to observe bubbling (low $\sigma$ combined with low $v$), while in the case of water/glycerol mixtures no bubbling could be observed (high $\sigma$). For these reasons as a third case, the alkane chemical group was selected for additional tests to further confirm our findings. Alkanes are acyclic saturated hydrocarbons with a chemical formula of $C_nH_{2n+2}$. In ambient conditions, they are in liquid phase for 4 < $n$ < 17. The chemical properties of the various molecules of this group are similar, but for increasing $n$, surface tension and viscosity increase.

Table 2 (in addition to data visualized in Figure 2) shows the outcomes for the different alkanes used in the experiments. It is possible to identify a transition regime between "break-up" and "surface levitation" regimes. Specifically, hexane and heptane (with $n = 6$ and $n = 7$ in Table 2) produce a break-up outcome, while alkanes with $n > 9$ were successfully levitated by the sublimating surfaces. The results confirmed the criteria established above: for hexane and heptane, both $\sigma$ and $v$ are below the critical values, $\sigma_C$ and $v_C$, respectively, and break-up regime is observed, whereas for other alkanes, for which both criteria of $\sigma > \sigma_C$ and $v > v_C$ are satisfied, the levitation regime is observed.

**Table 2.** Levitation outcome of various alkanes on the sublimating surfaces.

| $n$ | name | Surface tension (mN/m) | Kinematic Viscosity (mm$^2$/s) | Levitation Outcome |
|---|---|---|---|---|
| 6 | hexane | 18.40 | 0.47 | Break-up |
| 7 | heptane | 20.10 | 0.59 | Break-up |
| 10 | decane | 23.37 | 1.15 | Surface levitation |
| 12 | dodecane | 25.35 | 1.79 | Surface levitation |
| 13 | tridecane | 25.87 | 2.26 | Surface levitation |
| 14 | tetradecane | 26.56 | 2.78 | Surface levitation |
| 15 | pentadecane | 26.90 | 3.33 | Surface levitation |
| 16 | hexadecane | 27.47 | 4.01 | Surface levitation |

These results demonstrate that surface tension is a relevant liquid property because it counteracts the deformation of the vapor/liquid interface by the $CO_2$ vapor, attempting to penetrate in the liquid volume by forming bubbles. Also, kinematic viscosity is relevant to resist $CO_2$ vapor propagation inside the liquid drop and prevent the rise of vapor bubbles due to buoyancy. To provide an insight on the role of kinematic viscosity in preventing drop "break-up" behavior, we estimated the theoretical terminal velocities of $CO_2$ vapor bubbles of varying diameter, moving inside a viscous liquid. If we consider a $CO_2$ vapor bubble with diameter $\delta$, vapor density $\rho'$ and vapor dynamic viscosity $\mu'$ (in ambient conditions) moving inside a fluid medium freely under gravity, then its terminal velocity $V$ is given by the equation:[40]

$$V = \frac{1}{3}\frac{\delta^2 g}{\nu}\left(\frac{\rho'}{\rho} - 1\right)\frac{\mu + \mu'}{\mu + \frac{3}{2}\mu'} \tag{1}$$

where $g = 9.81$ m/s$^2$ (gravitational constant), $\rho$ and $\mu$ are the density and the dynamic viscosity, respectively, of the liquid in ambient conditions. The $CO_2$ vapor bubbles were assumed to be at room temperature and therefore the values of $\rho' = 1.842$ kg/m$^3$ and $\mu' = 0.0147$ mPa·s were used for the calculations.

Recent studies have proven theoretically that the shape of these bubbles can alter significantly the dynamics inside the liquid medium.[41] However, in the present case, as a first order analysis, we will assume that all bubbles have spherical shape, since the scope of this discussion is to show that liquid mediums of different kinematic viscosity can exhibit bubble terminal velocities of different orders of magnitude and this can affect significantly their stability under $CO_2$ sublimation. Calculations were performed for bubble diameter in the range 10 to 300 μm, with the upper value being one order of magnitude smaller than the tested drop size. In practice, an experimental estimation of the vapor bubbles size would be challenging, and requiring a high magnification observation apparatus. Figure 3 shows the theoretically estimated terminal velocities of $CO_2$ vapor bubbles moving in four liquids of different kinematic viscosities (0.59, 4.01, 100 and 1000 mm$^2$/s) as function of the bubble diameter, $\delta$. The red open circles correspond to heptane, exhibiting drop break-up. Heptane was chosen as a representative non-levitating liquid. Hexane and other liquids like silicone oils with lower kinematic viscosities would also give high bubble terminal velocities like heptane, but are not depicted here in order to make the graph less crowded and clearer. The other three more viscous liquids (hexadecane and higher viscosity silicone oils) were surface levitated (green curves in Figure 3). Clearly there is a strong correlation between bubble terminal velocity and liquid viscosity: small $CO_2$ vapor bubbles of diameter $\delta \approx 10$ μm move extremely slow in the viscous liquids (1000 mm$^2$/s has a bubble velocity $V \approx 3 \cdot 10^{-4}$ mm/s, 100 mm$^2$/s has $V \approx 3 \cdot 10^{-3}$ mm/s and 4.01 mm$^2$/s has $V \approx 8 \cdot 10^{-2}$ mm/s), with respect to the relevant time scales of the present experiment of transport and coalescence, which are in the order of tens or hundreds of ms, and the freezing event which is in the order of several seconds. These velocities are calculated assuming that the bubbles overcome the surface tension barrier and propagate inside the liquid volume. As shown in the graph of Figure 3, for all the liquid mediums, increasing their vapor bubble size ($\delta$) will increase the terminal velocities of the bubbles by orders of magnitude, but in the present work no optical analysis of the bubble size was performed to correlate it with drop "break-up", since this is beyond the scope. Therefore, it is expected that the bubble formation and transport process is very much resisted when the value of liquid kinematic viscosity is high, and that if a bubble does form e.g. due to the low values of $\sigma$, — it will only move slowly inside the liquid medium, with such a slow velocity as to not be observed by the present experimental setup, specifically in the present case where the observed phenomena occur in a very limited time scale. On the contrary, a $CO_2$ bubble of $\delta \approx 10$ μm which propagates in heptane will have a terminal velocity of $V \approx 666$ μm/s. If the size of the bubble increases up to $\delta \approx 300$ μm the terminal velocity will be $V \approx 6$ m/s.

Such terminal velocity values can definitely be observable within the time frame of the experiment and this explains why heptane shows "break-up" outcome.

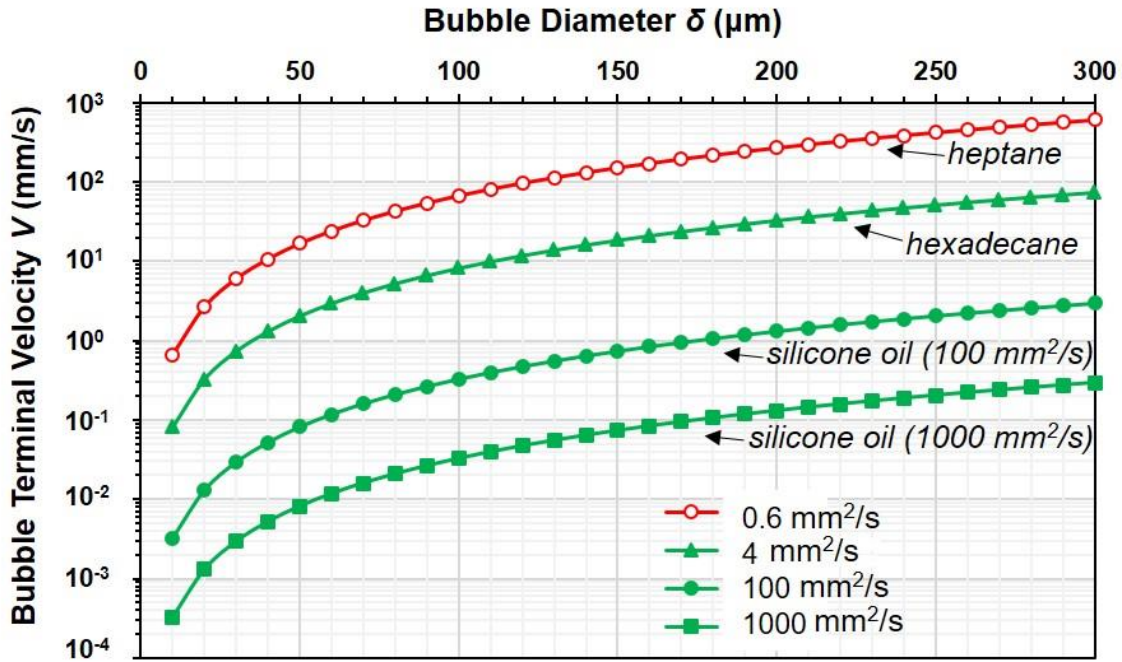

*Figure 3.* Terminal velocities (theoretical estimation from equation 1) of rising $CO_2$ bubbles inside a liquid as function of the drop diameter, Different curves correspond to liquids of different kinematic viscosity values, namely 0.59, 4.01, 100 and 1000 $mm^2/s$. Red color depicts the liquids that their contact with the sublimating $CO_2$ surface resulted in bubbling effect, while green color marks the ones that were surface levitated.

For comparison, one can also estimate the vapor velocity below the drop by simple scaling analysis. Assuming that the vapor layer below the drop has a cylindrical shape, with base diameter proportional to the drop diameter, $D_d$ (of the order of ~ 1 mm), and vapor layer thickness, $h$ (~100μm), the vapor velocity can be estimated as $V_v \propto \dot{m}_s / \rho_v D_d h$, where $\dot{m}_s$ is the vapor evaporation rate and has been already estimated to be of the order of $10^{-7}$ kg/s, and $\rho_v$ ~ 1 kg/m$^3$.[33] Using such scaling argument, the vapor velocity below the drop is of the order of 1 m/s, which is comparable to the speed of vapor bubbles of heptane as shown in Figure 3 for larger bubbles. Of course the aforementioned calculation is an order of magnitude estimate of bubble diameter and its terminal velocity. However, as a summary it is clear that $v$ is a parameter that influences significantly whether a liquid will be surface levitated or no. In addition, the aforementioned considerations indicate how low viscosity combined with low surface tension can promote drop break-up due to vapor flow within the liquid volume. Based on these values we can also provide an estimation of representative relevant non-dimensional numbers, such as the Eckert, Jakob and Lewis numbers. The Eckert number, which is mainly relevant to high speed flows where effects additional to convection and conduction may be important in the energy equation, represents the ratio of the kinetic energy of the $CO_2$ vapor flow and vapor enthalpy difference across the vapor layer thickness, and can be expressed as: $Ec = V_v^2/c_p \Delta T$, where $V_v^2 \approx 1$ m/s is the vapor speed (estimated above), $c_p \approx 800$ J/kg·K is the specific heat of the $CO_2$ and $\Delta T \approx 100$ K is the temperature difference across the vapor layer, i.e. between substrate and the drop. The resulting Eckert number value here is thus low, of the order of $10^{-5}$, and proves that for the present system, the energy balance is described solely by the standard conduction and convection mechanisms. However, from the fluid dynamical standpoint, the amount of released kinetic energy is sufficient to achieve drop levitation for most of the liquids while the high heat dissipation potential explains why the dry ice coatings are quite stable even in warm environments, way above their sublimation temperature.

The Jakob number (*Ja*), expressing the ratio between sensible and latent heat absorbed during solid-vapor phase change $Ja = c_p \Delta T/H_{vap}$, is of the order of $10^{-1}$ for the $CO_2$ vapor layer, since the latent heat of sublimation for $CO_2$ is $H_{vap} = 574$ kJ/kg. The only variable practically in *Ja* is the $\Delta T$, which could be controlled by changing the temperature of the liquid. However the operating range of temperatures is limited by the boiling and freezing points of the liquids and therefore no significant variations of *Ja* should be expected. Finally, the Lewis number $Le = \alpha/D$, which is the ratio of the thermal ($\alpha$) to mass ($D$) diffusivity for the $CO_2$ vapor layer, is $Le \approx 0.7$ (for ambient conditions $\alpha = 1.1 \cdot 10^{-5}$ m$^2$/s and $D = 1.6 \cdot 10^{-5}$ m$^2$/s). This *Le* value indicates that the thermal and mass diffusivities are of the same order of magnitude.

**Drop mixing/coalescence experiments.** After mapping the liquids that can be processed by surface levitation on a sublimating surface, drop mixing/coalescence experiments were performed on a sublimating coating, to show a proof-of-concept of liquid coalescence in a contactless manner on a sublimating platform. The idea is to continuously regenerate and sustain the sublimating coating on a cold aluminum substrate, by operating in a $CO_2$ vapor filled chamber (see Figure 4): as such, $CO_2$ below the transported drops can sublimate from the coating, to enable surface levitation, and subsequently de-sublimate from the environment to the coating, to regenerate it. The aluminum substrate was manufactured with open fluidic channels, to guide the drop transport (see Figure 5).

The concept is easily demonstrated by preliminary experiments performed combining glycerol drops (Movie M4), which can be difficult to process due to high adhesion to the substrate and high viscosity. In the Movie M4, showing a top view of the channel, it can be clearly seen that the glycerol drops, which remain in the levitation mode as demonstrated above (see also Figure 2), can roll off easily despite their high viscosity, and coalesce when their trajectories meet at the Y-junction. Note that the driving force for the motion is gravity, with the substrate being tilted by only few degrees from the horizontal position. The coalescence of the glycerol drops was performed with the V-shaped channels shown in Figure 5a. We should state here that when performing such coalescence experiments, depending on the liquid viscosity, solubility and reactivity, it is possible to achieve partial or full mixing, or chemical reactions that evolve with different time scales.[42-44]

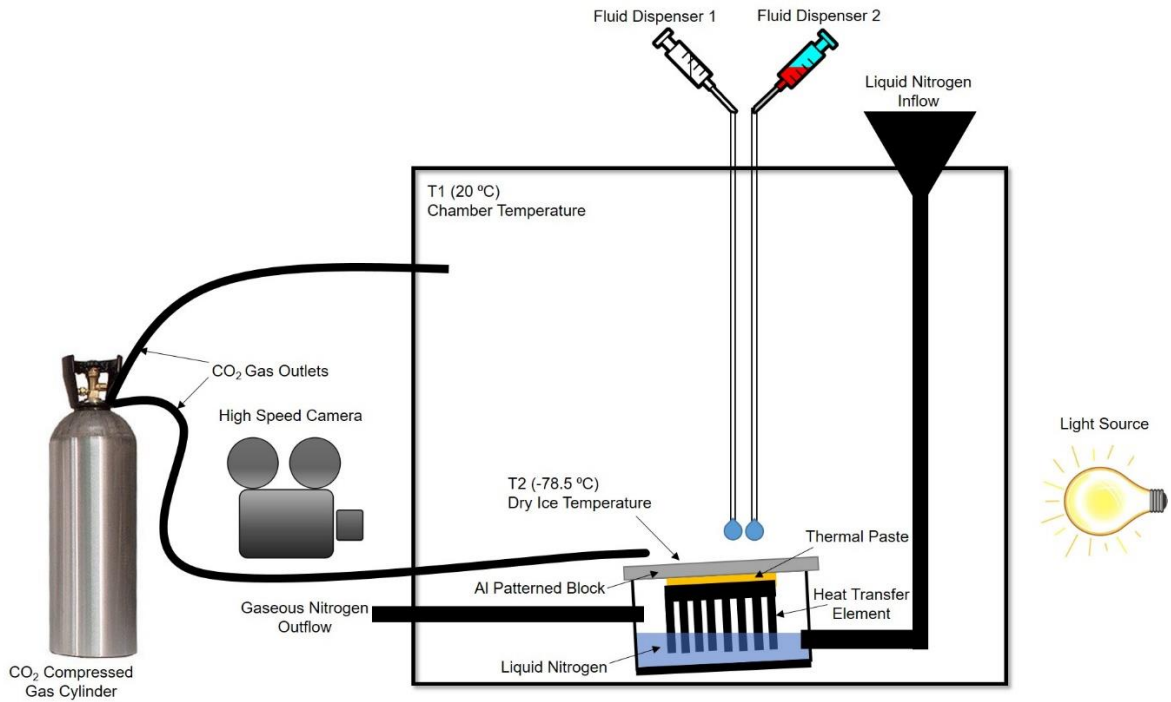

*Figure 4.* Sketch of the experimental setup for dry ice formation on the aluminum patterned channels.

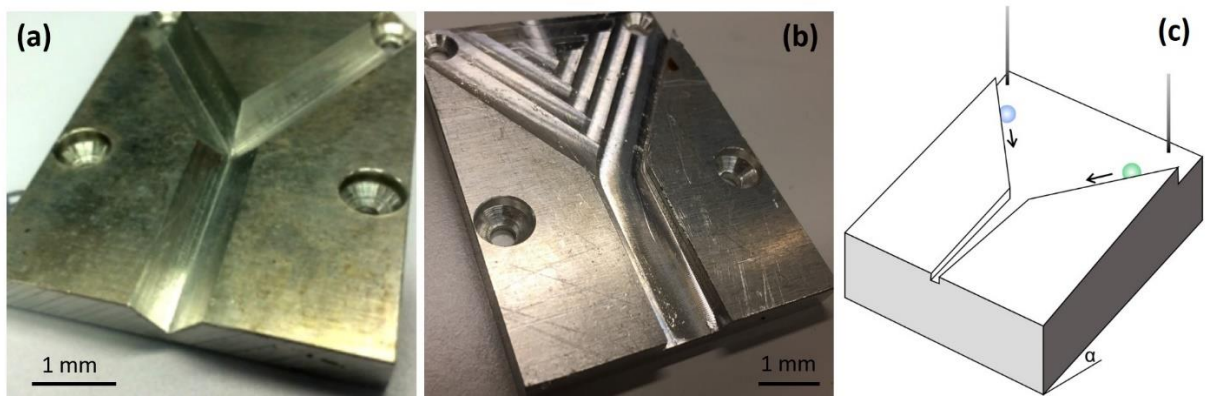

*Figure 5.* Aluminum substrate with channels for guiding the dispensed drops (a) with V-shape section and (b) with rectangular section. The aluminum substrates were coated by $CO_2$. (c) Schematic of the drop coalescence concept. A small tilt angle of the order of α = 2° is sufficient to facilitate the drop transport towards the coalescence region.

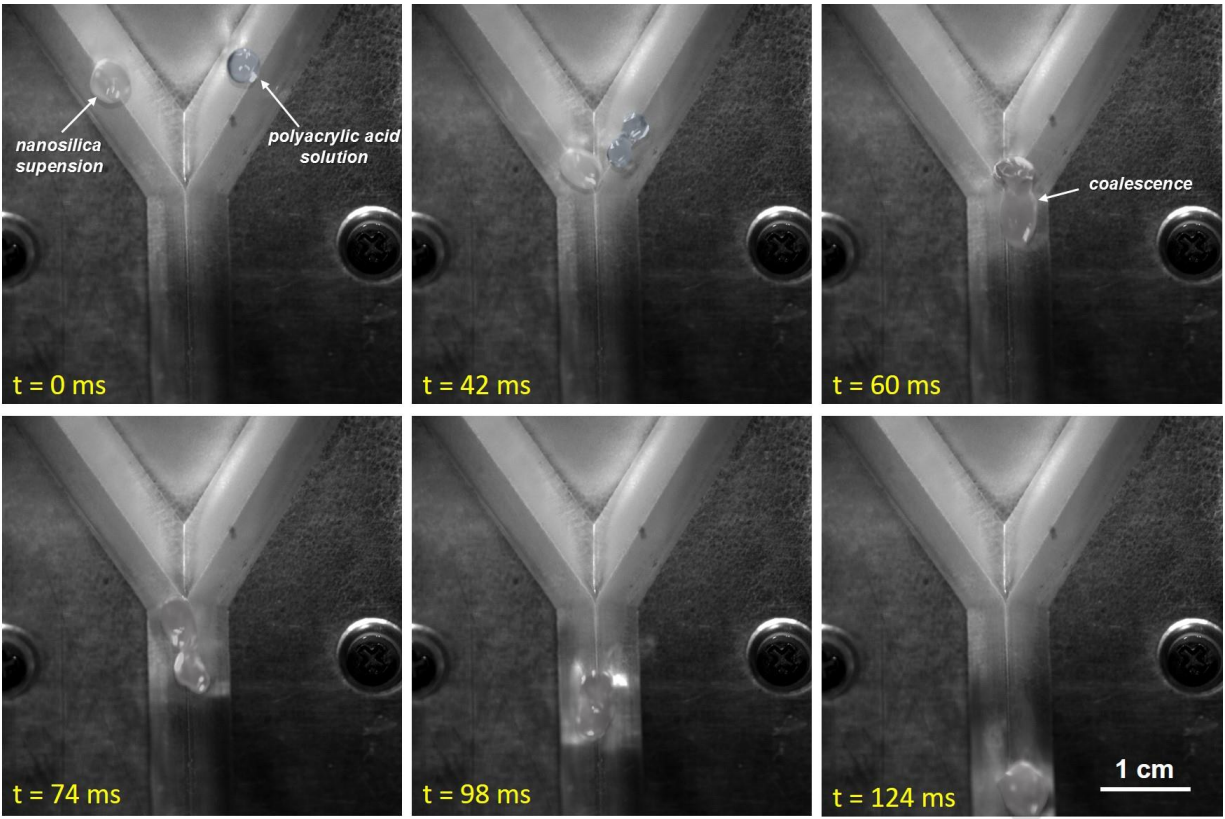

*Figure 6.* *Image sequence of the coalescence process of a 3 wt % nano-silica in water dispersion (white drop, left channel) and a 3 wt % polyacrylic acid dispersion in water (transparent drop, right channel) using the V-shaped channels. The drops have been slightly false colored to improve the image contrast for better visualization purposes. Full image sequence is available in Movie M5.*

To further prove the concept, additional tests were performed using various liquid-phase substances, including nanosuspensions, polymer solutions and non-Newtonian liquids, which customarily tend to adhere to and contaminate the surfaces/transport channels that are used for their processing. Here we show that such liquids can be easily processed in a contactless manner on the here presented apparatus. We choose to present here three representative examples to illustrate the concept: the first two demonstrate the coalescence between particle suspensions and solutions, and the third one demonstrates the coalescence between non-Newtonian liquids. It is worth noting that upon coalescence no satellite drop formation was observed, even in the case of low viscosity liquids, since the drops used throughout these experiments were of approximately equal size.[45] Moreover the formation of satellites drops due to impact upon dispensing was minimized by bringing the dispensing nozzles in close proximity with the dry ice coating to avoid splashing effects.[46]

The first coalescence example is illustrated by the image sequence in Figure 6, and the corresponding Movie M5, showing the coalescence process between a water-based nanoparticle dispersion and a water-based polymer solution with the channel pattern of Figure 5a (V-shaped channel). The nanoparticle suspension is composed of 3 wt % nanosilica powder (Aerosil 200 from Evonik Industries, Germany) in water and the polymer solution contains 3 wt % polyacrylic acid (Sigma) in water. Such particle suspensions are typically processed with conventional channels or tubes, or they are simply left to evaporate. However, the suspensions very easily tend to evaporate from most of the surfaces they come in contact with. In case of drop evaporation, the well-known "coffee ring effect" due to residue patterns left on a solid surface by a puddle of particle-laden liquid, can be typically observed.[47,48] With the present setup, using either nanoparticle suspensions or polymer solutions, we demonstrate that the two liquids can flow and coalesce without leaving any traces in the channels. As shown in Figure 6, the two drops start their rolling down motion in the channels after they are dispensed and they coalesce over a time scale of the order of 60 ms (frame 3). This demonstration shows that sublimating coatings could be used for preparing nanocomposite suspensions of small volumes, a process that could be particularly useful when continuous processing is required at the drop level, or the high material cost requires volume minimization. Along the channel where the nanosupension and the polymer solution were dispensed (upper left side) and inside the channel where coalescence occurs (center bottom), no sign of contamination was observed. The Al patterned channels were also inspected after the dry ice removal for any traces of contamination, but no stains could be identified after repeated experiments. Therefore we can conclude that an effective contactless coalescence occurs. Another demonstration of contactless mixing of an actual coffee drop with water was performed and the images sequences can be found in Figure S1 of the supporting information, while the Movie M6 shows the whole mixing process.

The second test case (Figure 7 and Movie M7) shows the coalescence procedure of a two-component silicone elastomer: part A and part B (CY52-276, Dow Corning). One of the two parts contains the base material (elastomer) and the other part contains the catalyst for the crosslinking reaction. When these two parts are mixed together, a catalyst based chemical reaction takes place and the final silicone rubber material is formed. These silicone elastomers exhibit unique viscoelastic properties because they are non-Newtonian liquids and they are widely used in soft molding because they can replicate all forms of structures that come to contact with nanoprecision.[49,50] The two components have a high dynamic viscosity of 1000 mPa·s, thus according to our results presented above (see Figure 2 and 3) they can be successfully surface levitated on a sublimating coating. Indeed, the image sequence in Figure 7 shows that the two components can successfully coalesce in a contactless manner. It is worth noting that these silicone-based materials exhibit very particular flow behavior and this can be noticed by the long tails that the drops form after they are dispensed. These tails stay connected with the dispensing nozzles but after a while, the tails are disconnected from the dispensing nozzles and they coalesce with the main drop (frame 4 to 5). Due to this unique rheological behavior, a large quantity of these materials is practically lost when they are processed with conventional materials, which makes the use of methods like the present desirable. As an additional note, we found that it was easier to process and transport the two components in the rectangular shaped channel (Figure 5b), rather than the V-shaped one (Figure 5a), since with this second shape the drops showed the tendency to partially stick to the narrow bottom part of channels, close to the sharp edge. The problem was eliminated by using the rectangular shaped channel.

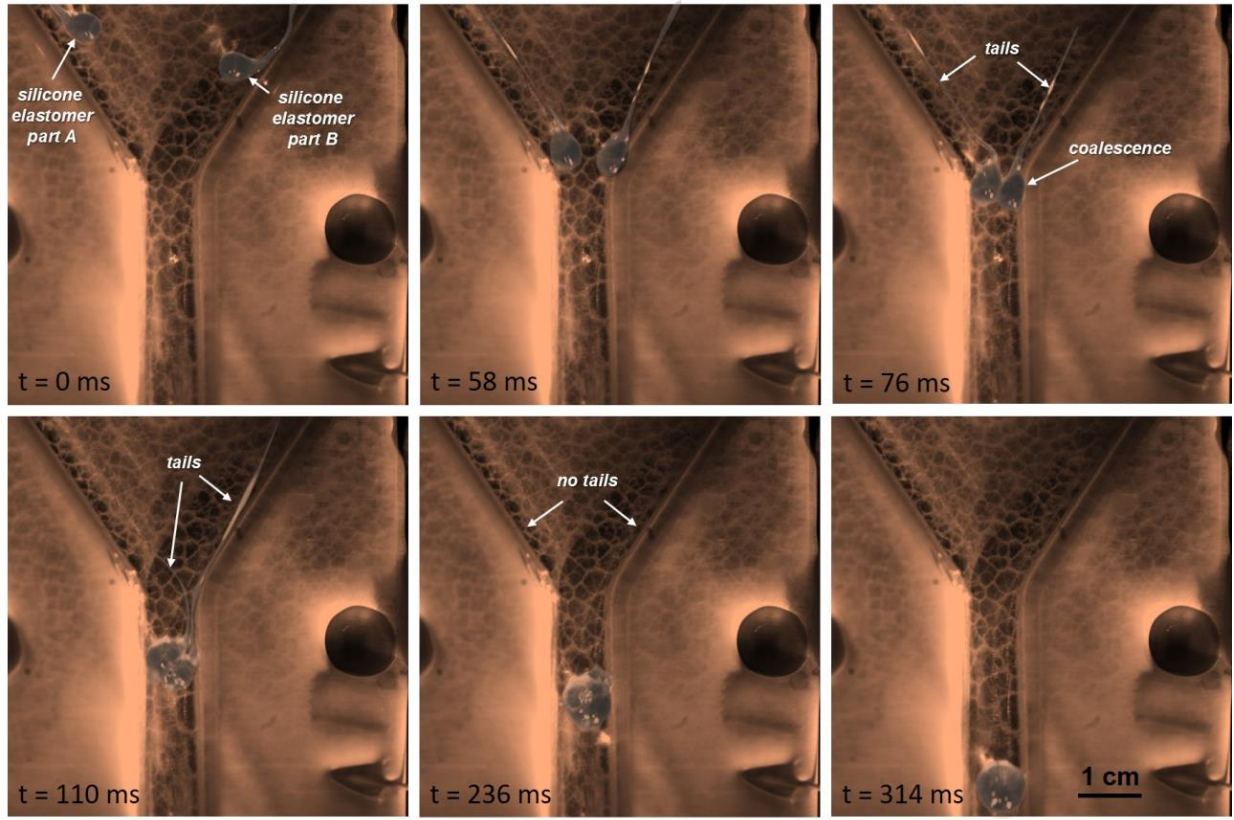

*Figure 7.* Image sequence of the coalescence process of a two-component silicone elastomer using the rectangular channel geometry (Figure 5b). The image has been slightly false colored for visualization purposes. Full image sequence is available in Movie M7.

## CONCLUSIONS

In this work we present a novel experimental method and platform to manipulate liquid drops by surface levitation and perform coalescence in a contactless and potentially continuous (sequence of drops) manner, taking advantage of self-sustained sublimating coatings. The $CO_2$ sublimating layer can be generated on demand inside an ad-hoc designed chamber, filled with $CO_2$ vapor, and can coat virtually any substrate, e.g., aluminum, as demonstrated in the present work.

To understand which liquids can be handled and processed, we first of all defined a map identifying the two possible regimes that can be observed on a sublimating surface: the "surface levitation" regime, and the "break-up" regime, in which bubbling of $CO_2$ vapor inside the drop is observed. On the basis of liquid surface tension and kinematic viscosity, we found that a liquid can experience surface levitation only if both liquid properties are higher than two critical values, i.e. $\sigma > \sigma_C \approx 20$ mN/m and $v > v_C \approx 0.9$ mm$^2$/s. The combined effect of surface tension and kinematic viscosity are extensively discussed in the manuscript and compared with theoretical estimates of bubble dynamics to explain the phenomenon. We then presented the proof-of-concept of a coalescence process on a self-sustained sublimating coating, showing representative tests with water-based nanodispersions, polymer solutions and non-Newtonian liquids. By using open channels having different section shape to guide the drops, we also analyzed the effect of the channel geometry when using different liquids. To conclude, we demonstrated the potential of processing drops on self-sustained sublimating surfaces, which can be beneficial to reduce significantly material loss and related contamination in various processes, and could find use in contactless transport of liquids in applications such as lab-on-a-chip, microfluidic and chemical processing devices.

**EXPERIMENTAL SECTION**

The schematic of the experimental setup used in this work is depicted schematically in Figure 4. The experiments were performed inside a closed cubic Plexiglas chamber, with an edge length of 25 cm. The purpose of the chamber is to maintain a pure $CO_2$ vapor environment, at 1 atm, with zero humidity. $CO_2$ is constantly supplied into the chamber through two plastic tubes that are connected to a $CO_2$ reservoir. The lower tube delivers the vapor close to the patterned aluminum substrate, where the $CO_2$ coating is intended to form. Once the dry ice coating is formed, the $CO_2$ supply from the lower tube is disconnected and the upper tube is used instead to maintain a slight overpressure in the chamber and a $CO_2$ environment, without affecting the drop transport on the sublimating surface. The thickness of the dry ice coating can be tuned by adjusting the flow of the $CO_2$ vapor in the chamber and the temperature of the aluminum substrate which affect the condensation rate. However, if the dry ice continuously builds up, a non-uniform coating of millimetric thickness will be obtained. In order to avoid this, we always maintained a visible, thin and smooth dry ice coating with thickness of the order of hundreds of microns. The patterned square aluminum surface contains the guiding channels, which form a Y shape to guide the drops to the coalescence point, has 5 cm side length and it is mounted on a thermally insulated cartridge containing the coolant fluid (liquid nitrogen). Two types of Y-shaped channels are used in this work as it is shown in Figure 5. The first aluminum substrate (Figure 5a) has a V-shaped section, with a sharp edge at the bottom, whereas the second substrate has a rectangular section (Figure 5b). For specific liquids, like silicone elastomers, it was found that the drop transport and coalescence was more effective with the second type of channel geometry (see Results and Discussion section for details). Liquid nitrogen is supplied via a thermally insulated tube until it partially fills the cartridge underneath the aluminum substrate. The aluminum substrate cooling is facilitated using a finned heat sink glued on its backside using a thermal paste. The heat transfer element is partially dipped in liquid nitrogen and this permits fast cooling of the aluminum substrate. The liquid nitrogen is continuously boiling inside the cartridge and an outlet tube allows for the nitrogen vapor that is formed to exit to the environment. To compensate for this loss of liquid nitrogen, there is constant supply from the cartridge inflow tube that compensates for the ongoing evaporation. During this process, the aluminum surface cools to approximately -79 ºC. The temperatures of the aluminum block and the chamber are continuously monitored by thermocouples via LabVIEW interface. When the aluminum substrate reaches the $CO_2$ phase change temperature ($\approx$ -79 ºC), the $CO_2$ vapor in the chamber starts to desublimate, thus leading to the formation of a solid $CO_2$ coating on the surface. Figure S2 shows representative image sequences of the coating formation process on a flat aluminum substrate (no channels). To minimize liquid nitrogen losses and ensure that dry ice will be formed only on the Al surface, the remaining external parts of the cartridge are thermally insulated. Two liquid dispensers are used to deliver the liquid drops to be coalesced via the Y-shaped, $CO_2$-coated channels. To optically monitor the dynamic phenomena that occur during the liquid coalescence, a Photron Fastcam PCI1024 high speed camera with a maximum resolution of 1024x1024 pixels was used, operating at 500, 1000 and 2000 frames per second. A halogen light source was used to illuminate the sample.

ASSOCIATED CONTENT

**Supporting Information**.

Figure S1. Image sequence of the coalescence process of a coffee and a water drop using the V-shaped channels; Figure S2. Image sequence of $CO_2$ coating formation on a flat aluminum substrate; Movie M1. A silicone oil drop (surface tension 20.6 mN/m and kinematic viscosity 20 mm$^2$/s) is deposited on a $CO_2$ sublimating surface; the drop is in surface levitation regime; Movie M2. A perfluorohexane drop (surface tension 11.9 mN/m and kinematic viscosity 0.4 mm$^2$/s) is deposited on a $CO_2$ sublimating surface. The drop is in break-up regime; Movie M3. An isopropyl alcohol drop deposited on the sublimating surface. Notice that the same bubbling behavior of the initial drop applied also in the two satellite droplets generated later (10$^{th}$ s in the movie); Movie M4. Coalescence of two pure glycerol drops at 2000 fps (replay is 67x slower); Movie M5. Coalescence of a 3 wt. % silica nanosuspension in water with a 3 wt. % polyacrylic acid solution in water at 500 fps (replay is 17x slower); Movie M6. Coalescence of a coffee and a pure water drop at 500 fps (replay is 17x slower); Movie M7. Coalescence of part A and part B of a silicone elastomer kit at 500 fps (replay is 17x slower).

AUTHOR INFORMATION

**Corresponding Author**

*E-mail: dpoulikakos@ethz.ch

**Author Contributions**

The manuscript was written through contributions of all authors. All authors have given approval to the final version of the manuscript. ‡These authors contributed equally.

**Funding Sources**

We acknowledge partial support of the Swiss National Science Foundation under Grant 162565 and the European Research Council under Advanced Grant 669908 (INTICE). CA acknowledges funding from European Community through a Marie Curie Intra-European Fellowship (ICE$^2$, 301174).

**Notes**

The authors declare no competing financial interest.

**ACKNOWLEDGMENT**

We would like to thank Mr. Jovo Vidic from the Laboratory of Thermodynamics in Emerging Technologies of ETH Zurich for his contribution in the construction of the experimental setup and Mr. Emmanuel Heer from the Department of Mechanical and Process Engineering of ETH Zurich for his help in the data acquisition.

**Graphical abstract**

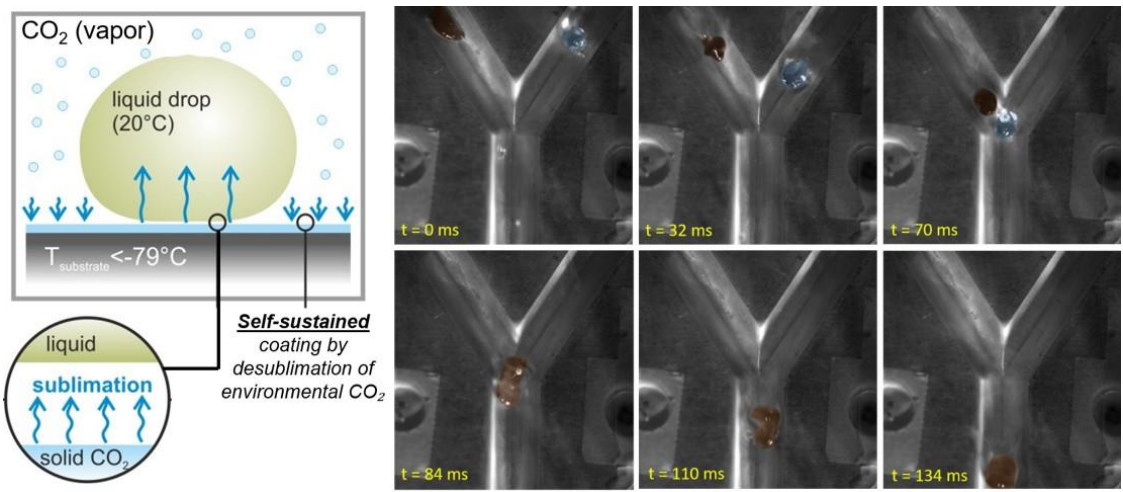